# Controlling the thermal contact resistance of a carbon nanotube heat spreader

Kamal H. Baloch

Institute of Physical Science and Technology & Department of Materials Science and Engineering, University of Maryland, College Park, MD 20740 (USA)

Norvik Voskanian, John Cumings [a]

Department of Materials Science and Engineering, University of Maryland, College Park, MD 20740 (USA)

**The ability to tune the thermal resistance of carbon nanotube mechanical supports from insulating to conducting could permit the next generation of thermal management devices. Here, we demonstrate fabrication techniques for carbon nanotube supports that realize either weak or strong thermal coupling, selectively. Direct imaging by *in-situ* electron thermal microscopy shows that the thermal contact resistance of a nanotube weakly-coupled to its support is greater than 250 K·m/W and that this value can be reduced to $4.2^{+5.6}_{-2.1}$ K·m/W by imbedding the nanotube in metal contacts.**

[a] Corresponding Author, email: cumings@umd.edu

Due to their unique thermal properties, [1-4] Carbon nanotubes (CNTs) have generated interest in the scientific community. Even though many questions relating to their electrical and mechanical properties have been resolved, they remain elusive in their thermal properties. Evidence indeed suggests them to be superior thermal conductors, but the reported thermal conductivity values vary widely between 200 and 3000 W/K·m in the literature. [2, 5-12] This incongruity in some cases can be attributed to variations in thermal contact resistance $R_c$, [13] a parameter that has recently received much attention, with studies on the subject reporting values that vary widely. [13, 14-22] For example, Maune *et al.* perform thermometry employing electrical breakdown of CNTs to report $R_c$ values of 0.6--3.0 K·m/W, [15] whereas Tsen *et al.* report a value for $R_c$ of 25 K·m/W from photothermal current microscopy relying on local heating from a scanning laser. [17] Unfortunately, in both cases the heat sources are not amenable to independent characterization and the underlying phenomena thus still leave substantial uncertainties in $R_c$ for the two studies. In this work, we rely upon a well-characterized metallic palladium (Pd) heater wire as a power source to heat a CNT. Using this, we demonstrate that $R_c$ can be manipulated by more than two orders of magnitude, making it possible to realize thermally conducting or thermally insulating contacts as may be needed for different applications. Reproducibility and control of $R_c$ would open the door for the next generation of nanoscale thermal management and thermal logic devices such as thermal diodes and thermal transistors. [23, 24]

To perform these studies, we employ electron thermal microscopy (EThM) described elsewhere. [25] In EThM, thermal images are obtained by transmission electron microscope (TEM) observations of the in-situ solid to liquid phase transition of low-melting-point (156.6 °C) nanoscale indium (In) islands, acting as binary thermometers, evaporated on the back of electron transparent silicon nitride ($SiN_x$) substrates.[26] In the present study, measurements are done on two types of samples. High resolution TEM images of the two types of devices prior to In deposition are shown in figure 1a and figure 1b. In one type (figure 1a), a CNT (diameter ~14 nm ), thermally anchored only by the SiN substrate beneath, is heated by a Pd metal contact connected to a resistive Pd heater wire. In another sample type (figure 1b), a second Pd contact is added to the free end of another CNT (diameter ~17 nm) to control the contact resistance with the substrate. For clarity, a schematic top view of the latter device is depicted in Figure 1c. A schematic cross-section of the anchored region of CNT is shown in Figure 1d. Note that the CNT only contacts the $SiN_x$ substrate underneath tangentially, while the width of the overlap between the CNT and Pd metal above is nearly half the circumference of the CNT. This difference in contact area is the primary mechanism we report for tuning the thermal contact resistance. The details of the fabrication processes of these devices are explained in the Supplemental Material. In the absence of a nanotube, the distance from the center of the Pd heater wire to which islands melt on either side is expected to be the same at a given voltage, reflecting the geometrical symmetry of the structure. However, the presence of a nanotube should introduce an asymmetry in this distance due to its high thermal conductivity (approximately 1000--3000 W/K·m, refs. 6-10) relative to the substrate (~4 W/K·m, ref. 25), and the amount of the asymmetry can produce a reliable measure of the contact resistance, $^{SiN}R_c$, between the nanotube and the substrate. In the case of the device shown in Figure 1a, the thermal conductivity of the region on the right of the heater should be higher than of that on the left, showing melting at lower applied voltages. However, thermal maps obtained using EThM (shown in Figure 2a with a bright-field TEM image overlay) show no such asymmetry. This indicates that even though the CNT has a high thermal conductivity, its ability to transport heat to the substrate is limited by a high thermal contact resistance $^{SiN}R_c$. Other experiments on two similar devices also show this lack of asymmetry, demonstrating that a CNT adhered to a $SiN_x$ substrate has an inherently high $^{SiN}R_c$.

Figure 2b shows the distance from the center of the heater wire to selected In islands plotted versus the voltages required to melt them, together with a fit from linear regression (details are given in the Supplemental Material). Also shown on the same plot is a second data set extracted from indium islands at a symmetric point on the opposite side of the heater wire. By comparison, it is readily apparent that there is no significant difference in melting distance on the two sides of the heater wire, indicating a high thermal contact resistance between the nanotube and the substrate.

Theoretically-predicted thermal maps are generated using finite-element modeling, with results shown in figure 2c. From our observations, we can conclude that the asymmetry is smaller than a minimum-detectable level and that the $^{SiN}R_c$ value is therefore higher than a corresponding minimum value. The details of the finite-element modeling are given in the Supplemental Material and the $^{SiN}R_c$ value that we extract is 250 K·m/W. However, it is important to note that this value is a lower bound for $^{SiN}R_c$, and the actual value could be much higher than this. This is a high thermal contact resistance, exceeding other published values, and it indicates that even though the nanotube and substrate are in intimate contact, there is effectively good thermal insulation between them.

These results are apparently not consistent with other studies, which show a smaller contact resistance for a CNT adhered to a substrate. [13, 15-20] However, many of these other studies use SEM imaging, [13, 16, 19, 20] which is known to deposit significant amounts of carbonaceous material onto CNTs during imaging. [27] This material may reduce $R_c$ in the same manner that Pd does here. Similarly, other studies use a CNT manipulation-and-placement routine that explicitly calls for embedding the CNT in an acrylic adhesive. [2, 28] For other studies that do not use SEM imaging or other adhesive coatings, [15, 17, 18] the measurements are performed at elevated temperatures, where near-field thermal radiation [29] may be playing a significant role in the transport phenomena, reducing the observed $R_c$ value. The results we report here do not use SEM imaging or adhesives, are well-characterized by TEM to exclude the possible presence of gross amounts of contaminating materials, and are carried out at relatively low temperatures, where thermal radiation effects are expected to play a diminished role in the transport. Thus, many possible sources of uncertainty are mitigated in the studies we present here.

One remaining alternative explanation for this reduced asymmetry rather than a high $R_c$ value might be a diminished thermal conductivity of the nanotube, as might be expected for example from electron-beam induced damage.[30] However, other studies show that nanotubes retain their high thermal conductivities under similar temperatures and electron microscopy imaging conditions.[31] To further demonstrate that nanotubes retain high thermal conductivity in our experimental setup, we also present results from a second device designed to reduce the observed $R_c$ value and provide a test of the thermal conductivity of the nanotube. This is achieved by adding a second Pd thermal pad at the opposite end of the CNT as shown in Figures 1b and 1c. The Pd thermal pad located at the center of the heater wire facilitates the transport of heat into the CNT, while the other Pd thermal pad on the right side of the heater wire helps the CNT dispense heat more effectively into the substrate beneath.

Figure 2d shows a thermal map of this second device, analogous to Figure 2a. Here, the middle of the heater wire is the hottest region, but unlike the un-anchored CNT case, a clear asymmetry can be seen in the thermal contours. The area around the Pd thermal pad to the right of the heater melts the In islands at significantly lower voltages than an equidistant area on the left side, where there is no CNT. The relationship between melting voltage and distance from the heat source for a set of In islands near the Pd anchor and a symmetric point on the opposite side of the heater wire are plotted in Figure 2e, where a clear difference can be seen between the two sides. As described in the Supplemental Material, we performed linear regression and finite-element modeling to extract an expectation value and 95% confidence interval for $^{Pd}R_c$ of $4.2^{+5.6}_{-2.1}$ K·m/W. Figure 2f shows a voltage map from the modeling, showing an asymmetry comparable to the experimental voltage map. This low value for $^{Pd}R_c$ by itself demonstrates that the nanotube must have high thermal conductivity and thus is not affected significantly by beam damage. It is important to mention here that all data were acquired under imaging conditions such that the heating due to the electron beam was not a factor, as described in the Supplemental Material.

The observation that the asymmetric melting of the In islands occurs only when two Pd thermal contacts are deposited on either end of the CNT and the fact that the $^{SiN}R_c$ and $^{Pd}R_c$ differ by more than an order of magnitude both support our assertion that manipulating the effective contact width can substantially change the thermal contact resistance. If we assume a model in which $R_c$ is inversely proportional to the contact width, we expect that the higher contact area of CNT anchored to the Pd metal would give lower thermal resistance than when it is lying on the substrate, as indicated in the schematic in Figure 1d. For multiwall CNTs, there are inherent difficulties in defining a contact width between the CNT and the substrate. In fact, mechanical modeling predicts almost no flattening for a CNT of the sizes we report here, when van der Waals bonded to a flat substrate.[32] The exact contact area of the nanotube and the $SiN_x$ would require considerations of the surface roughness of the substrate and the mechanical interactions of the nanotube, which are outside the scope of the present work. However, we note that the membranes have very low surface roughness (< 8 Å RMS) and the basic result is the same even when the the experiment is repeated with the nanotube on the back side of the membrane, which is expected to have an even lower roughness, comparable to the parent Si wafer on which it was grown. It is also instructive to compare the thermal conductivity and electrical conductivity of the CNT-Pd contacts to determine whether the enhancement of thermal transport may be electron-mediated. In electrical devices fabricated with similar geometries, we routinely see an electrical contact resistance of approximately 10 k$\Omega$, consistent with other findings. [33] Using the Wiedemann-Franz law, we thus estimate a thermal conductance from electrons of $10^{-9}$ W/K. This value is two orders of magnitude smaller than our modeled thermal conductance of $1.2 \times 10^{-7}$ W/K at the contacts. Thus the heat transfer between the CNT and Pd is believed to be phonon-mediated.

In summary, we demonstrate that the thermal coupling between a CNT and its mechanical support can be manipulated. In fact, through this study we have shown that controlling the thermal nature of mechanical supports of a MWCNT should no longer be considered a limiting factor in utilizing CNTs for thermal applications. This result should aid in the future engineering of CNTs for thermal devices.

We acknowledge the support of the Maryland NanoCenter and its NispLab. The NispLab is supported in part by the NSF as a MRSEC Shared Experimental Facility. This work has been supported by the U.S. Nuclear Regulatory Commission under grant NRC3809950 and by the UMD-NSF-MRSEC under grant DMR 05-20471. We also acknowledge support from an Agilent Technologies University Research Grant.

**REFERENCES**

[1] J. Hone, B. Batlogg, Z. Benes, A. T. Johnson, and J. E. Fischer. Science 289, 1730 (2000)

[2] P. Kim, L. Shi, A. Majumdar, and P. L. McEuen. Rev. Lett., 215502 (2001)

[3] M. J. Biercuk, M.C. Llaguno, M. Radosavljevic, J.K. Hyun, A. T. Johnson, and J.E. Fischer. Appl. Phys. Lett. 80, 2767–2769 (2002)

[4] J. Hone, M.C Llaguno, M.J. Biercuk, A.T. Johnson, B. Batlogg, Z. Benes, J.E. Fischer. *Appl. Phys. A* 74, 339-343 (2002)

[5] S. Berber, Y. K. Kwon, and D. Tomanek. Phys. Rev. Lett. 84, 4613 (2000)

[6] E. Pop, D. Mann, Q. Wang, K. Goodson, and H. J. Dai. Nano Lett. 6, 96 (2006)

[7] S. Maruyama. Physica B 323,193 2002.

[8] D. J. Yang, Q. Zhang, G. Chen, S. F. Yoon, J. Ahn, S. G. Wang, Q. Zhou, Q. Wang and J. Q. Li. Phys. Rev. B 66, 165440 (2002)

[9] T .Y. Choi, D. Poulikakos, J. Tharian, and U. Sennhauser. NanoLett. 6, 1589 (2006)

[10] M. Fujii, X. Zhang, H. Q. Xie, H. Ago, K. Takahashi, T. Ikuta, H. Abe, and T. Shimizu. Phys. Rev. Lett. 95, 065502 (2005)

[11] C. H. Yu, L. Shi, Z. Yao, D. Y. Li, and A. Majumdar. Nanotube NanoLett. 5, 1842 (2005)

[12] J. P. Small, K. M. Perez, and P. Kim. Phys. Rev. Lett. 91, 256801 (2003)

13. I. K. Hsu, R. Kumar, A. Bushmaker, S. B. Cronin, M. T. Pettes, L. Shi, T. Brintlinger, M. S. Fuhrer, and J. Cumings. Appl. Phys. Lett. 92, 063119 (2008)

14. L. Shi. App. Phys. Lett. 92, 206103 (2008)

15. H. Maune, H. Y. Chiu, and M. Bockrath. Appl. Phys. Lett. 89, 013109 (2006)

16. C. H. Yu, S. K. Saha, J. H. Zhou, L. Shi, A. M. Cassell, B. A. Cruden, Q. Ngo, and J. Li. J. Heat Transfer 128, 234-239 (2006)

17. A. W. Tsen, L. A. K. Donev, H. Kurt, L. H. Herman and J. Park. Nature Nanotechnology 4, 108 - 113 (2008)

18. E. Pop; D. A. Mann, K. E. Goodson, H. Dai. J. of Appl. Phys. 101, 093710 (2007)

19. P. Kim, L. Shi, A. Majumdar, P. L. McEuen. Physica B 323, 67-70, (2002)

20. L. Shi, J. Zhuo, P. Kim, A. Batchtold, A. Majumdar, P. L. McEuen. J. Appl. Phys. 105, 104306 (2009)

21. Pettes, M. T.; Shi, L. Adv. Funct. Mater. (2009), 19, 3918-3925

22. R. S. Prasher, X. J. Hu, Y. Chalopin, N. Mingo, K. Lofgreen, S. Volz, F. Cleri, and Pawel Keblinski. Phy. Rev. Lett. 102, 105901 (2009)

23. B. Li, L. Wang and G. Casati. Thermal. Phys. Rev. Lett.93, 184301 (2004)

24. B. Li, L. Wang and G. Casati Appl. Phys. Lett. 88, 143501 (2006)

25. T. Brintlinger, Y. Qi, K.H. Baloch, D. Goldhaber-Gordon, and J. Cumings. Electron Thermal Microscopy. Nano Lett. 8, 2, 582-585 (2008)

26. SiNx membranes obtained from Silson Inc

27. M. F. Yu, O. Lourie, M. J. Dyer, K. Moloni, T. F. Kelly, and R. S. Rouff. Science 287, 5453 (2000)

28. H. Dai, J. H. Hafner, A. G. Rinzler, D. T. Colbert, and R. E. Smalley. Nature 384, 147 (1996)

29. S. Shen, A. Narayanaswamy, and Gang Chen. Nano Lett. 9, 8, (2009)

30. N. G. Chopra, L. X. Benedict, V. H. Crespi, M. L. Cohen, S. G. Louie and A. Zettl. Nature 377, 135-138 (1995)

[31] G. E. Begtrup, K. K. Ray, B. M. Kessler, T. D. Yuzvinsky, H. Garcia and A. Zettl. Phys. Rev. Lett. 99, 155901 (2007)

[32] T. Hertel, R. E. Walkup and P. Avouris. Phys. Rev. B, 58, 20 (1998)

[33] M. S. Purewal, B. H. Hong, A. Ravi, B. Chandra, J. Hone, and P. Kim. Phys. Rev. Lett. 98, 186808 (2007)

**FIGURE CAPTIONS**

**FIGURE 1 (a)** and **(b)** TEM images of the thermally anchored and unanchored CNT devices used for this study, prior to depositing In with both scale bars of 1 μm. **(c)** Schematic diagram of thermally anchored device with a circuit overlay. **(d)** Schematic cross-section of the sample, showing that the CNT touches the SiN$_x$ membrane only tangentially whereas the contact of the CNT with the Pd contact width is approximately half the circumference.

**FIGURE 2 (a)** and **(d)** Experimental thermal maps with the device overlay obtained by assigning a unique color to the voltage required to melt each island. The scale bars are 1 μm. White lines are added equidistant from the center of the heater as guides to the eye. The thermal map in **(d)** shows the asymmetric melting of In islands on the two sides of the heater wire whereas in **(a)**, no such asymmetry can be readily detected. **(b)** and **(e)** are the plots of average distances (with standard deviation) at which the In islands melt for each given voltage, for the two devices. The data in **(b)** show no clear asymmetry while data in **(e)** show a clear asymmetry in the melting of islands at each voltage, quantitatively. **(c)** and **(f)** Simulated thermal maps obtained using finite element analysis.

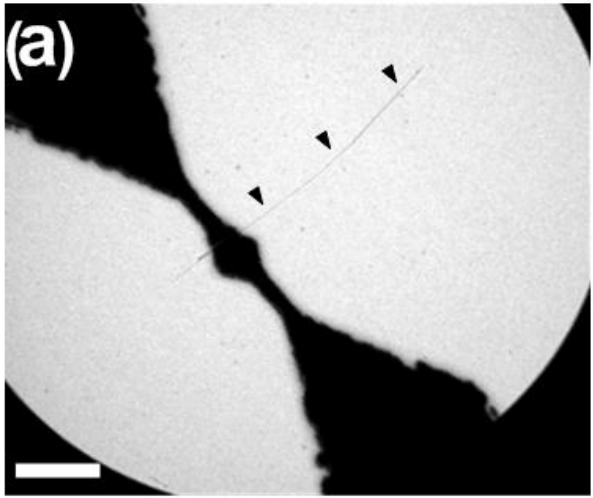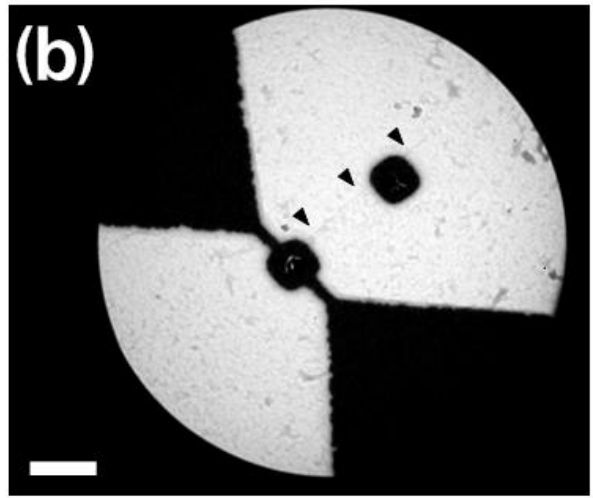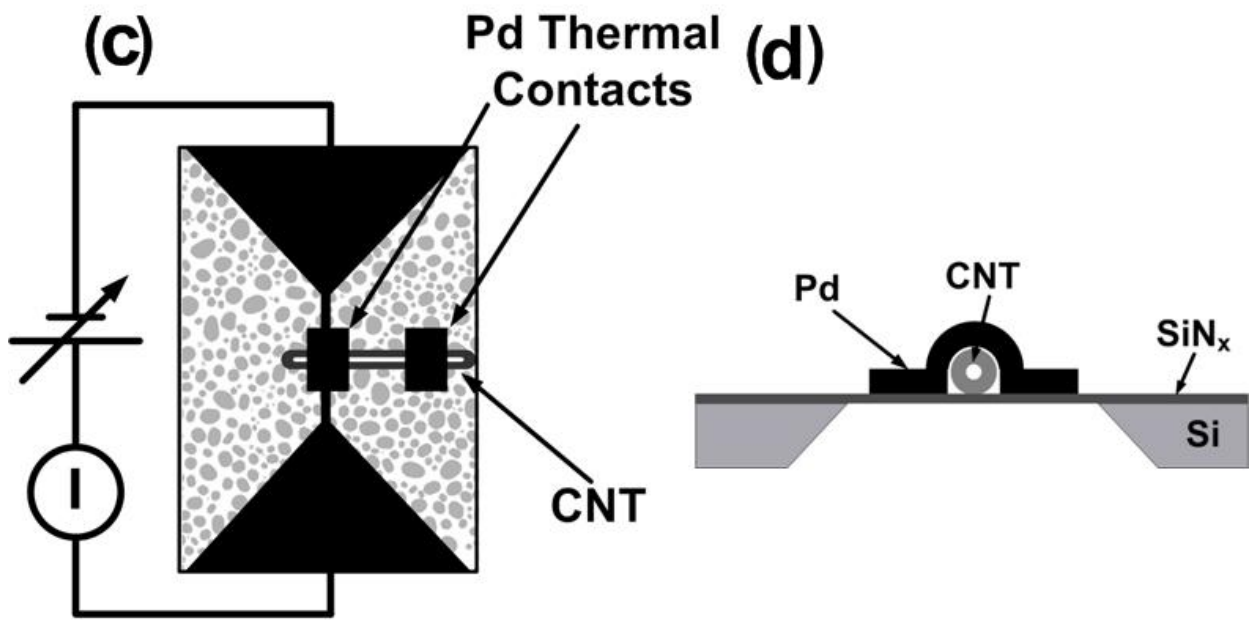

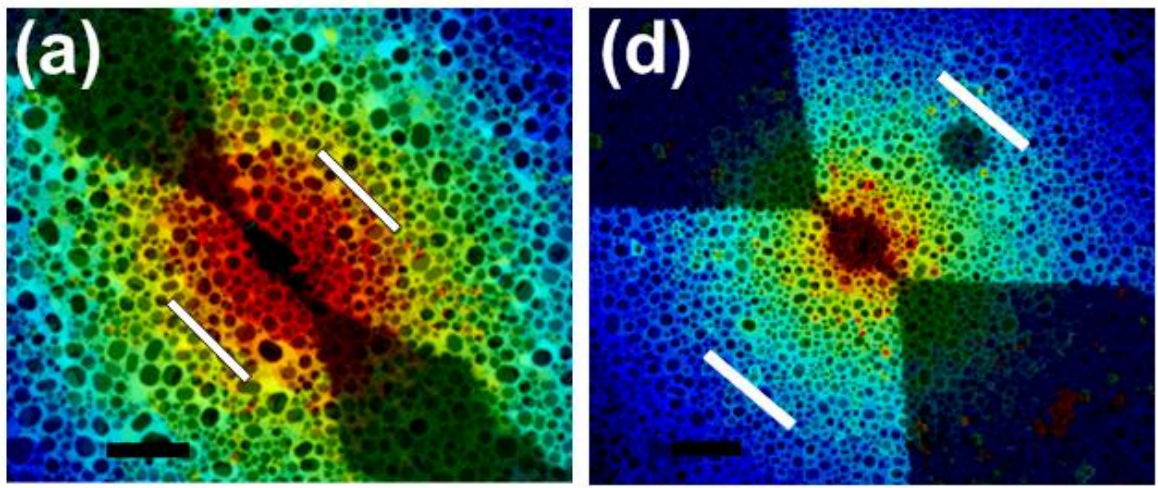

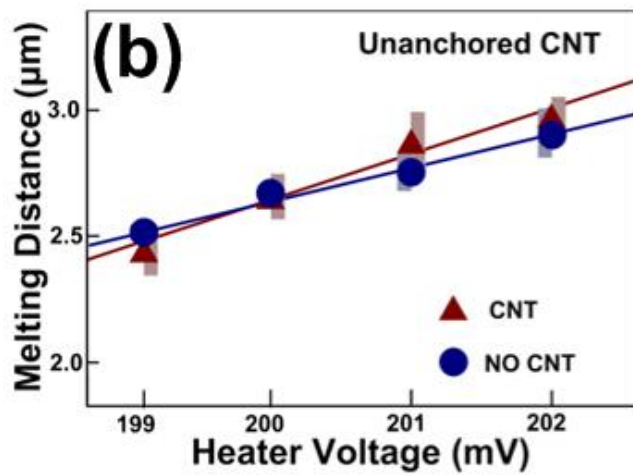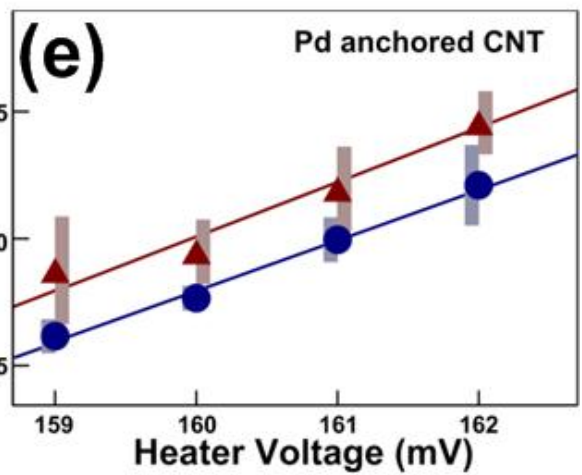

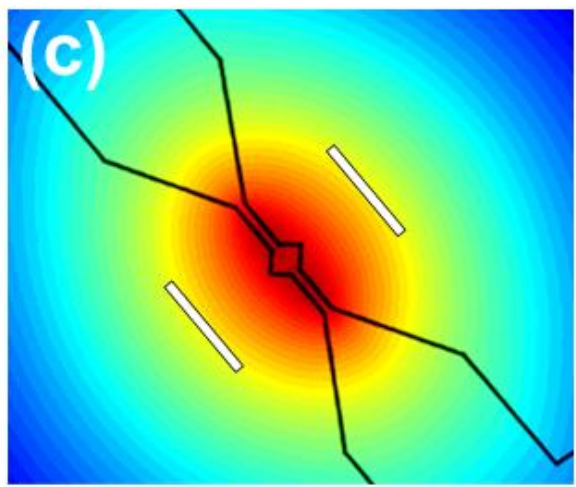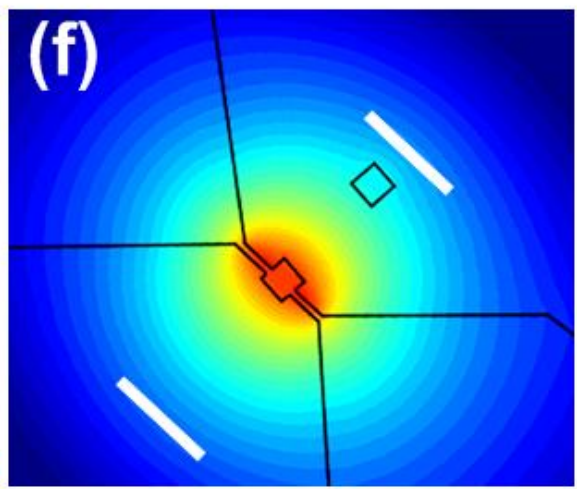

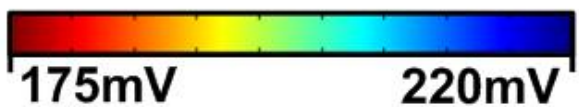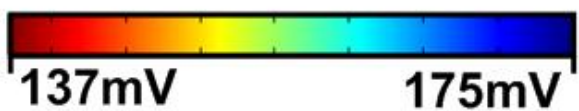

## SUPPLEMENTAL MATERIAL

1. **Sample Preparation**

   Devices for this study are fabricated in multiple steps on commercially available, free-standing and electron-transparent $SiN_x$ membranes. High quality arc-discharge grown multiwall CNTs [S1] are spin-cast onto the substrate, and subsequently Pd thermal contact pads with attached Pd heater wires are added by conventional electron beam lithography. Finally, a thin layer of In is deposited on the back side of the $SiN_x$ membrane, which forms a discontinuous layer of isolated islands. A high resolution TEM image of this device prior to In deposition is shown in figure 1a. The additional device fabricated by incorporating a second Pd thermal contact pad at the other end of the CNT is shown in figure 1b. EThM studies allow us to make clear conclusions about the $R_c$ value for the nanotube on the bare substrate ($^{SiN}R_c$) as well as for $R_c$ for the nanotube underneath the Pd thermal contact ($^{Pd}R_c$).

2. **Finite element modeling**

   Modeling of the data is performed in two dimensions using the finite element analysis package Comsol. Combined electrical and thermal partial differential equations are used in an iterative solver. In the model, the working thermal equation is

   $$\nabla \cdot (K \nabla T) - \frac{\Delta T}{R_c^{2D}} + P = 0 \qquad (S1)$$

   where $T$ is the local temperature, $\Delta T$ is the temperature difference between the CNT and the substrate, $K$ is thermal conductivity, $R_c^{2D}$ is the two dimensional thermal contact resistance of the CNT with the material it is contact with, and $P$ is the power in the heater wire, given by $P = \sigma |\nabla V|^2$. In this relationship V is the electric potential, and $\sigma = \sigma_o (1 + \alpha \Delta T)^{-1}$, is the electrical conductivity, with $\sigma_o$ the value at room temperature and $\alpha$ the temperature coefficient, both characterized previously.[25] In the modeling, steady state conditions were assumed. Since the power source and thermal conductivity values are previously characterized, the only unknown parameter in equation (S1) is $R_c^{2D}$. A series of simulated thermal maps are obtained by choosing $R_c^{2D}$ as a free parameter. The contact resistance is extracted from the simulated thermal maps that match with the one obtained experimentally, and variations in $R_c^{2D}$ are obtained from repeated simulations using the various voltage values for melting the indium obtained as described below. Finally, $R_c^{2D}$ values are converted to one-dimensional $R_c$ values reported in the text by dividing by the width of the nanotube in the finite element model. In each case, a meshing study was performed to ensure accurate simulations, independent of computational finite element mesh spacing, within 1% of reported values. In our modeling, we assume that the thermal contact between the Pd and $SiN_X$ is perfect. Such interfaces are also known to exhibit thermal boundary resistance, but even in the worst case of two dissimilar materials, [S2] the temperature drop at this interface would be less than 10% of total temperature drop between the CNT and substrate.

3. **Method to obtain experimental melting voltages**

To measure $^{SiN}R_c$ quantitatively, we extract the voltage required to produce melting at a point 2/3 along the nanotube and at a symmetric point on the opposite side of the heater wire. These locations were chosen as they show the largest temperature-asymmetry in the finite-element model and are thus the most sensitive to variations in $^{SiN}R_c$. In quantitative analysis, we note that the In islands used to measure temperature come in a distribution of sizes, and confinement effects[S3] therefore produce an uncertainty in the melting temperature of any given island. If we use just a single In island on each side to detect a voltage asymmetry, this results in a large uncertainty in the amount of asymmetry in the temperature profiles, adding error to our measurement of $^{SiN}R_c$. To improve our estimate of $^{SiN}R_c$, we instead use an ensemble of islands to extract a more precise measure of the asymmetry. This is done by selecting sets of at least 30 In islands each in proximity to the nanotube as described below and calculating an average melting voltage and confidence interval for each set. To accomplish this, a control region is defined within a given radius around the point of interest and an identical symmetric region is defined on the opposite side of the heater wire. The radius of these regions is chosen so that they both contain at least 30 islands. For all islands within these two regions, the melting voltage and the distance from the center of the wire are extracted and plotted as shown in figure 2b of the text. Then these values are fit using a linear regression algorithm to obtain the expectation value and 95% confidence interval for the voltage needed to bring the substrate to the melting point of indium at the control point and also at the symmetric point. These expectation and 95% confidence interval values are used in the finite-element modeling described above to extract the thermal contact resistance between the CNT and substrate. We can estimate the smallest $^{SiN}R_c$ value that could generate an asymmetry consistent with our uncertainties by using a worst-case combination of these expectation values and confidence intervals and varying $^{SiN}R_c$ as a free parameter to obtain the same asymmetry within the theoretical model. Using a best-case combination of melting voltages would produce an upper bound on $^{SiN}R_c$, but such a combination produces a negative asymmetry and would thus predict and unphysical negative $^{SiN}R_c$. Thus, we can only meaningfully extract a lower bound on $^{SiN}R_c$.

To measure $^{Pd}R_c$, we repeat the process described above using EThM images for the nanotube with two Pd metal contacts, with the following modifications. Here, the control points are the center of the second Pd contact and a symmetric point on the other side of the heater wire, and when extracting $^{Pd}R_c$, we have the ability to calculate upper and lower bounds explicitly, as given in the text.

4. **Beam heating**

Independent measurements using a specimen holder with a calibrated heat source and thermocouple confirm that beam heating at high illumination intensity can increase the sample temperature by tens of °C, but the beam conditions used here produce heating that is immeasurably small within the resolution of the heating holder, namely less than 1°C.

**REFERENCES FOR SUPPLEMENTAL MATERIAL**

[S1] Arc discharge grown multiwalled carbon nanotubes obtained from Sigma-Aldrich

[S2] H.-K. Lyeo and D. G. Cahill. Thermal conductance of interfaces between highly dissimilar materials. Phys. Rev. B 73, 144301, (2006)

[S3] G. L. Allen, R. A. Bayles, W. W. Gile, and W. A. Jesser. Thin Solid Films 144 (1986)